\documentclass{emulateapj}

\slugcomment{Submitted: \today}
\shorttitle{Kes~17}
\shortauthors{Lee et al.}

\begin{document}

\def\kms{km~s$^{-1}$}
\def\cms{cm$^{-3}$}
\def\vecJ{{\bf J}}
\def\mum{$\mu$m}
\def\simlt{\lower.5ex\hbox{$\; \buildrel < \over \sim \;$}}
\def\simgt{\lower.5ex\hbox{$\; \buildrel > \over \sim \;$}}
\def\gt{>}
\def\lt{<}
\def\hh{\rm H$_2$}
\def\hone{\ion{H}{1}}

\title{Far-Infrared Luminous Supernova Remnant Kes~17}

\author{Ho-Gyu Lee\altaffilmark{1},
Dae-Sik Moon\altaffilmark{1}, 
Bon-Chul Koo\altaffilmark{2},
Takashi Onaka\altaffilmark{3}, 
Woong-Seob Jeong\altaffilmark{4},
Jong-Ho Shinn\altaffilmark{4},
and Itsuki Sakon\altaffilmark{3}
}

\altaffiltext{1}{Department of Astronomy and Astrophysics, 
University of Toronto, Toronto, ON M5S 3H4, Canada; 
hglee@astro.utoronto.ca,
moon@astro.utoronto.ca}
\altaffiltext{2}{Department of Physics and Astronomy, 
Seoul National University, Seoul 151-742, Korea; 
koo@astrohi.snu.ac.kr}
\altaffiltext{3}{Department of Astronomy, Graduate School of Science,
The University of Tokyo, Bunkyo-ku, Tokyo 113-0033, Japan; 
onaka@astron.s.u-tokyo.ac.jp,
isakon@astron.s.u-tokyo.ac.jp}
\altaffiltext{4}{Korea Astronomy and Space Science Institute, 
776, Daedeok-daero, Yuseong-gu, Daejeon 305-348, Korea;
jeongws@kasi.re.kr, 
jhshinn@kasi.re.kr}

\begin{abstract}
We present the results of infrared (IR; 2.5--160 $\mu$m) observations of the
supernova remnant (SNR) Kes~17 based on the data obtained with the $AKARI$ 
and $Spitzer$ satellites.
We first detect bright continuum emission of its western shell in the mid- and far-IR wavebands
together with its near-IR molecular line emission.
We also detect hidden mid-IR emission of its southern shell after subtraction 
of the background emission in this region.
The far-IR luminosity of the western shell is $\sim$ 8100 $L_{\rm \odot}$,
which makes Kes~17 one of the few SNRs of significant far-IR emission.
The fittings of the spectral energy distribution indicate the existence of two dust components:
$\sim$ 79~K (hot) and $\sim$ 27~K (cold)
corresponding to the dust mass of 
$\sim$ 6.2~$\times$~10$^{-4}$ $M_\odot$ and $\sim$ 6.7 $M_\odot$, respectively.
We suggest that the hot component represents the dust emission of the material 
swept up by the SNR to its western and southern boundaries,
compatible with the distribution of radio continuum emission overlapping the mid-IR emission 
in the western and southern shells.
The existence of hot ($\sim$ 2,000~K), shocked dense molecular gas revealed by the near-IR molecular line
emission in the western shell, on the other hand, suggests that the cold dust component
represents the dust emission related to the interaction between the SNR and nearby molecular gas.
The excitation conditions of the molecular gas appear to be consistent with those from
shocked, clumpy admixture gas of different temperatures.
We discuss three possibilities for the origin of the bright far-IR emission 
of the cold dust in the western shell:
the emission of dust in the inter-clump medium of shocked molecular clouds,
the emission of dust in evaporating flows of molecular clouds engulfed by hot gas,
and the emission of dust of nearby molecular clouds illuminated by radiative shocks.
\end{abstract}


\section{Introduction}

The broad-band infrared (IR) observations are well suited for the study of the evolution of SNRs. 
Besides the general advantage of suffering less extinction effects, 
there are diverse important phenomena related to supernova remnants (SNRs) which are bright in the IR wavebands.
In the near- and mid-IR regimes, for example, one can study the distributions of 
dense ejecta from core-collapse supernova explosions \citep[e.g.,][]{koo07, moo09},
circumstellar material produced by mass loss of the progenitors \citep[e.g.,][]{dwe08, lee09g39, lee09g292},
shock interactions between SNRs and nearby molecular clouds \citep[e.g.,][]{bur88, oli90, rea05, neu07, hew09, shi09},
and hot dust synthesized in ejecta \citep[e.g.,][]{are99, rho08},
along with the emission from the central compact 
objects \citep[e.g.,][]{moo04, kap06, wan06}.

The far-IR observations of SNRs, on the other hand, are relatively rare but 
very efficient to investigate cold dust which often contributes to a significant 
portion of the total mass of the dust in SNRs. 
First, it is studied that the swept-up dust in SNRs is responsible for 
the most of cooling of SNRs and plays a critical role in the evolutions of SNRs 
in dense medium \citep{dpsr87}.
Secondly, the far-IR observations of SNRs appear to provide better means of studying the dust 
formed during the process of supernova explosion \citep[e.g.,][]{sib10, bar10}.
Finally, if an SNR is interacting with a nearby molecular cloud, 
its far-IR emission is expected to be significantly enhanced by
the swept-up, inter-clump medium in the cloud or by the evaporation of 
dense clumps \citep[e.g.,][]{dra81, dwe81},
although the reality of these processes is yet to be confirmed observationally.

Kes 17, known as G304.6$+$0.1, is such an object in which broad-band IR 
observations can play an important role in our understanding of SNRs. 
Until recently this SNR has only been studied in radio continuum  
and OH line observations \citep{sha70, mil75, whi96, fra96},
revealing broken radio shell structures in the west and south.
The recent $Spitzer$ observations, however, detected very bright IR emission in 
the 3--8~$\mu$m range \citep{lee05, rea06}, which makes it in fact one of the
brightest SNRs in the wavelength range.
The follow-up spectroscopic observations in the 5--30~$\mu$m range
also detected many strong \hh\ emission lines \citep{hew09}.
Very recently, the XMM-Newton X-ray observations detected centrally filled thermal 
X-ray emission inside the non-thermal radio shell, listing it as a member 
of the class of mixed-morphology SNRs \citep{com10}.
All these indicate that Kes~17 is interacting with nearby molecular gas and 
that more IR observations are needed to better understand its nature.
In this paper, we present extensive broad-band IR observations of Kes~17
using the $AKARI$ satellite, together with new analyses of archival 
$Spitzer$ and radio continuum data.
For the distance to Kes~17, only a lower limit of 8~kpc,
corrected for the distance between the Sun and the Galactic center of 8.5 kpc,
is known from the observations of \hone\ line 
\citep{cas75}.
We adopt a distance of 8~$d_8$~kpc where $d_8$ is a scaling factor in this paper.

\section{Observations and Data}

\subsection{\textit{AKARI} observations}

The $AKARI$ imaging observations of Kes~17 were carried out
covering the mid-IR (13--27 $\mu$m) and far-IR (50--180 $\mu$m)
bands on 2007 February 5 and 6, respectively.
The journal of the $AKARI$ observations is given in Table~\ref{tab_obssum}.
The mid-IR observations were made by the 
Infrared Camera (IRC) equipped with an Si:As detector array
of 256 $\times$ 256 pixels which 
produced two images of 10\arcmin\ $\times$ 10\arcmin\ 
centered on 15 $\mu$m (IRC L15) and 24 $\mu$m (IRC L24) \citep{ona07}.
The total on-source integration time was 196~s for both images.
The basic calibration and data handling such as
dark subtraction, linearity fitting, distortion correction,
flat fielding, image combination and astrometric measurement 
were performed by IRC Imaging Data Reduction Pipeline version
20110225\footnote{http://www.ir.isas.jaxa.jp/ASTRO-F/Observation/DataReduction/IRC/}.
The far-IR observations, on the other hand, were made by the Far-Infrared Surveyor (FIS)
in two round-trip scans in the cross-scan shift mode \citep{kaw07}.
The scan speed and shift length were 15\arcsec\ s$^{-1}$ and 240\arcsec, respectively,
and the resulting imaging size was 40\arcmin\ $\times$ 12\arcmin\ elongated in the scan direction.
All the four FIS band (N60, Wide-S, Wide-L, and N160; 
see Table~\ref{tab_obsimg} for the details of the bands) 
images were obtained simultaneously in a single observing run
using Ge:Ga (20 $\times$ 2 pixels for N60; 20 $\times$ 3 pixels for Wide-S)
and stressed Ge:Ga (15 $\times$ 3 pixels for Wide-L; 15 $\times$ 2 pixels for N160) detector arrays.
The initial calibration and data handling such as 
glitch detection, dark subtraction, flat fielding and flux calibration
were performed by FIS Slow-Scan Toolkit version
20070914\footnote{http://www.ir.isas.jaxa.jp/ASTRO-F/Observation/DataReduction/FIS/},
followed by advanced image construction based on the refined sampling mechanism.
A summary of imaging bands used in this paper,
including six $AKARI$ bands (IRC L15, IRC L24, FIS N60, FIS Wide-S, FIS Wide-L, and FIS N160)
and three $Spitzer$ bands (IRAC 4.5, IRAC 8, and MIPS 24),
is given in Table~\ref{tab_obsimg}.

The $AKARI$ spectroscopic observations were carried out 
in the near-IR slit mode on 2008 August 8 and 2009 February 4--6
(Table~\ref{tab_obssum}), which produced grism spectra of R $\simeq$ 120 
spectral resolving power in the 2.5--5.0 $\mu$m wavelength range \citep{ohy07}.
Spectra of the two bright peaks and the narrow filament in the western shell of Kes~17
previously identified in $Spitzer$ IRAC images \citep{lee05, rea06} were obtained.
The slit positions and their coordinates are presented in Figure~\ref{fig_ir} and 
in Table~\ref{tab_slit}, respectively.
The 5\arcsec\ $\times$ 0\arcsec.8 slit was used for the target observations, whereas
background spectra, which were subtracted from the target spectra later, 
were simultaneously obtained using the 3\arcsec\ $\times$ 1\arcsec\ 
slit positioned 1\arcmin\ apart from the targets. 
The data calibration and handing were performed by IRC Spectroscopy Toolkit version
20081015\footnote{http://www.ir.isas.jaxa.jp/ASTRO-F/Observation/DataReduction/IRC/}.

\subsection{\textit{Spitzer} Infrared and \textit{ATCA} Radio Continuum Data}

Previous $Spitzer$ studies detected bright IR emission of Kes~17 in the IRAC bands
between 3.6 and 8.0 $\mu$m bands \citep{lee05, rea06}.
We used the $Spitzer$ IRAC images of 4.5 and 8.0 $\mu$m bands in this paper.
We analyzed the $Spitzer$ archival data
of MIPS 24 $\mu$m band\footnote{ 
http://data.spitzer.caltech.edu/popular/mipsgal/20080718\_enhanced/MIPS\_24/}
\citep{car09}
to make an image of $\sim1^\circ$ size of Kes~17. 
For the 24 $\mu$m band image, we used the $Spitzer$ data in this paper,
unless explicitly mentioned otherwise. 
In addition, we obtained 
the $Spitzer$ IRS 8--38 $\mu$m and MIPS SED 52--97 $\mu$m spectra 
available from the $Spitzer$ data
archive\footnote{http://sha.ipac.caltech.edu/applications/Spitzer/SHA/}
to estimate contributions of line emission in the wavelength ranges.

We also obtained a high-resolution (9\arcsec.17 $\times$ 7\arcsec.36)
radio-continuum image of Kes~17 using the Australia Telescope Compact Array ($ATCA$) archival data provided by 
the Australia Telescope Online Archive.\footnote{http://atoa.atnf.csiro.au}.
We synthesized the radio-continuum image of Kes~17
from the 12-h exposure 1.4~GHz data obtained with the 1.5A array configuration 
on 2004 March 24 after calibration of the data with 1329$-$665.

\section{Infrared Morphology of Kes~17}

\begin{figure}
\plotone{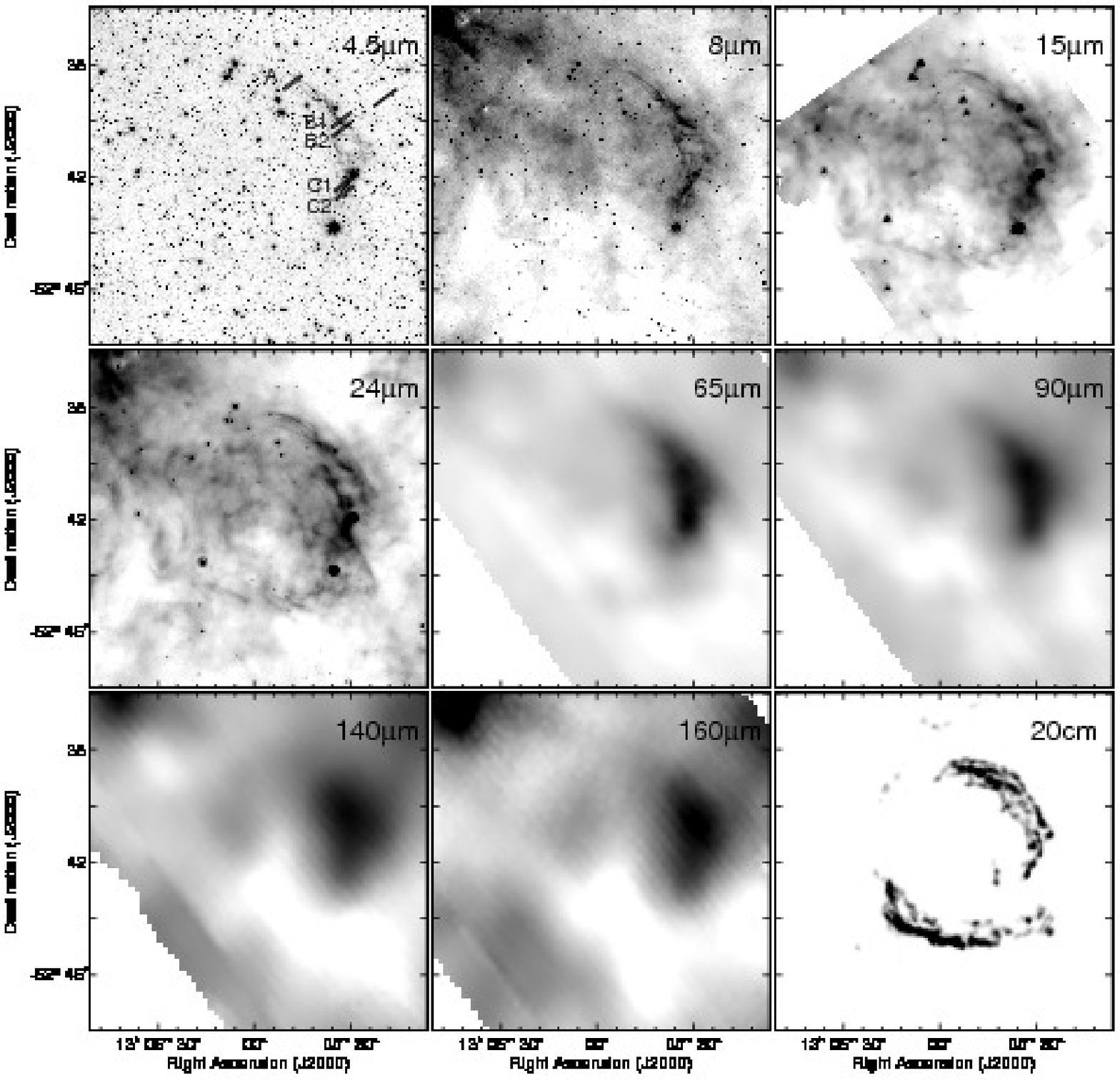}
\caption{$AKARI$ and $Spiter$ IR images of Kes~17, together with the $ATCA$ 20~cm image.
The wavelength of each imaging band is inserted at the upper-right corner of each panel.
The 4.5, 8, and 24 $\mu$m images are from $Spitzer$, and the rest IR images are from $AKARI$.
The slit positions of near-IR spectroscopic observations are shown
in the 4.5 $\mu$m image (see Table~\ref{tab_slit} for the coordinates).
The ranges of the gray scales are
1--10, 36--58, 43--56, 32--45, 160--290, 400--570, 1050--1300, and 700--850 MJy sr$^{-1}$
for the 4.5, 8, 15, 24, 65, 90, 140, and 160 $\mu$m band images, respectively.
The range of the gray scale of the radio continuum image is 0.001--0.01 Jy beam$^{-1}$.
}
\label{fig_ir}
\end{figure}

Figure~\ref{fig_ir} presents the $AKARI$ and $Spitzer$ near- to far-IR 
(4.5--160 $\mu$m) band images of Kes~17, together with the $ATCA$ 20~cm radio-continuum image.
The most conspicuous feature in the IR images is the bright emission of $\sim$ 5\arcmin\
in the western shell apparent in all of the IR bands.
The emission appears to be clumpy and filamentary in the near- and mid-IR bands up to 24 $\mu$m, 
whereas it becomes unresolved in the longer (65--160 $\mu$m) wavelength images  
due to increasingly large beam sizes.
This IR emission in the western shell partly overlaps that of
the radio continuum emission.
In addition to the bright western shell,
there is relatively weak mid-IR (15 and 24 $\mu$m) emission
in the southern shell overlapping the radio continuum emission in this region.

\begin{figure}
\plotone{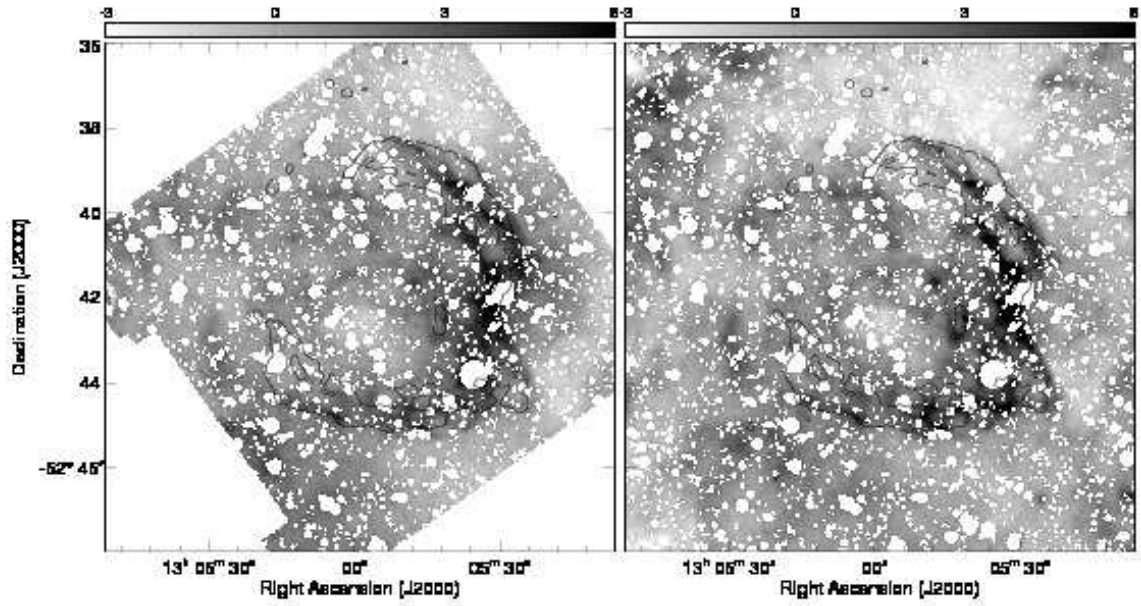}
\caption{The background-removed 15 (left) and
24 $\mu$m (right) band images of Kes~17.
The superposed contour is  
the 20~cm radio continuum emission boundary of 0.002 Jy beam$^{-1}$.
}
\label{fig_mid}
\end{figure}

There exists noticeable background emission in the direction of Kes~17
in all the images longer than 4.5 $\mu$m (Figure~\ref{fig_ir}), especially in the northeast.
In order to identify the southern shell emission in the 15 and 24 $\mu$m images
more clearly, we subtract the background emission from both the images as follows:
first, given that both the 4.5 and 8 $\mu$m emission of Kes~17 is due to the
same origin (i.e., emission from shocked \hh; see \S~4 for the details)
in the western shell, we obtain a linear correlation with a correlation coefficient $r$ $\simeq$ 0.86
between the 4.5 and 8 $\mu$m emission of the shell.
Based on the correlation, we scale the 4.5 $\mu$m emission to 8 $\mu$m and subtract
the scaled emission from the observed 8 $\mu$m emission to create a background image at 8 $\mu$m.
Next, we obtain a linear correlation of $r$ $\simeq$ 0.89 between the background 8 $\mu$m emission and 
the 24 $\mu$m emission outside the boundaries of Kes~17, and then scale the 8 $\mu$m emission to 24 $\mu$m
using the correlation.
We subtract the scaled emission from the observed emission at 24 $\mu$m,
which finally gives the background-removed 24 $\mu$m image of Kes~17 (Figure~\ref{fig_mid}).
The peak surface brightnesses of the western shell after the background subtraction are
6.1~$\pm$~1.5 and 6.4~$\pm$~1.5 MJy sr$^{-1}$ for 15 and 24 $\mu$m bands, respectively.
We apply the same method to obtain the background-removed $AKARI$ 15 and 24 $\mu$m images.
In all these processes, we convolved the images of different beam sizes with
that of the 24 $\mu$m images.
In Figure~\ref{fig_mid} we can now easily identify the existence of the southern shell 
at 15 and 24 $\mu$m which overlaps that of the radio continuum.
For far-IR images of 65, 90, 140, and 160 $\mu$m,
we estimate the background emission by taking the median value in the region between
4$'$ and 6$'$ radius from the center of Kes 17.
After background subtraction, the peak surface brightnesses of the western shell are
108~$\pm$~24, 135~$\pm$~45, 173~$\pm$~84, and 117~$\pm$~55 MJy sr$^{-1}$ 
for 65, 90, 140 and 160 $\mu$m bands, respectively. 

\begin{figure}
\plotone{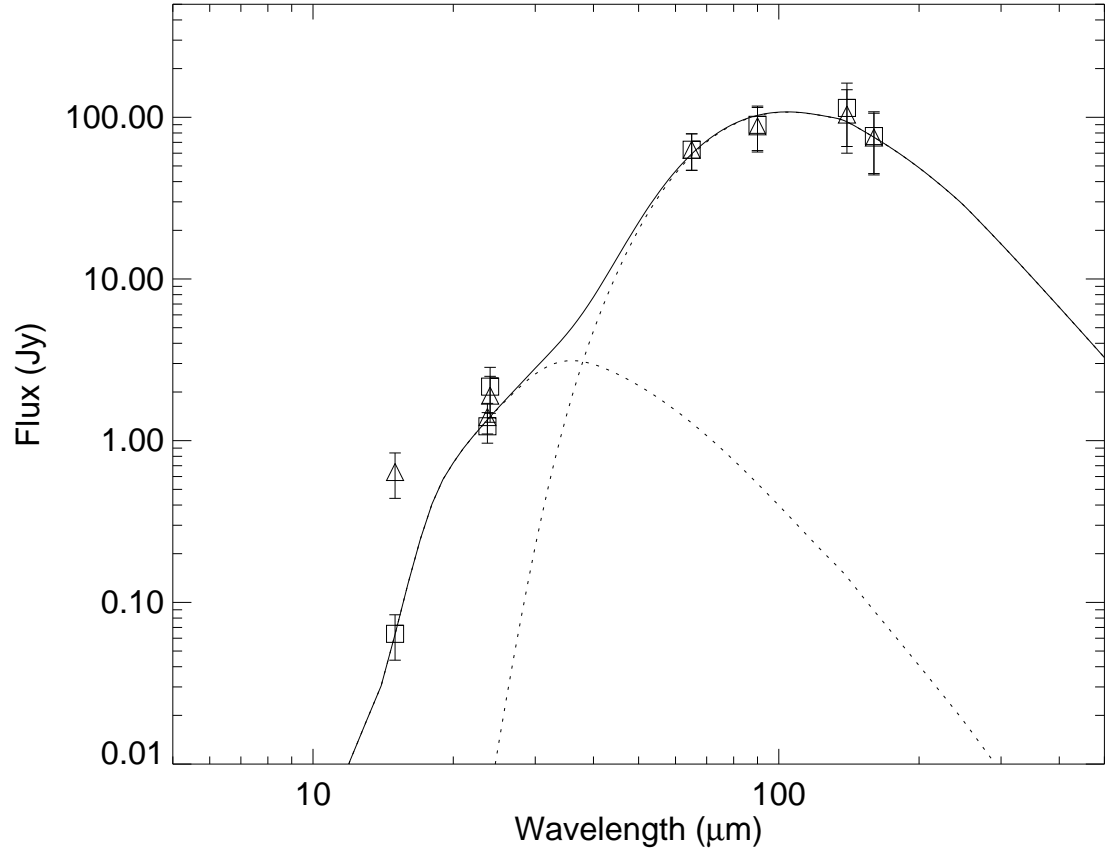}
\caption{The IR fluxes of the western shell of Kes~17:
the triangles are measured fluxes; the squares are color-corrected fluxes.
We use the line-subtracted fluxes except 140 and 160 $\mu$m band fluxes
where the line contributions are insignificant (see \S~3).
The two dotted curves show the modified blackbody radiation of
79~$\pm$~6~K and 27~$\pm$~3~K. 
The solid curve shows the combination of the two dotted curves.
}
\label{fig_sed}
\end{figure}

Table~\ref{tab_flux} presents the measured (or upper limits) IR fluxes of 
the western and southern shells between 15 and 160 $\mu$m,
together with the estimation of contributions of line emission in the bands.
The errors in the flux measurements are mainly caused by the uncertainties in 
the background measurements and also by those associated with the calibrations of 
the $AKARI$ and $Spitzer$ observations.
We estimate the line contributions to the observed fluxes of the 15 and 24 $\mu$m bands
using the $Spitzer$ IRS 8--38 $\mu$m spectra.
We construct the IRS spectra by subtracting the background obtained 
outside the shell structure and integrate them over the bandpass.
The estimated line contributions are 75~$\pm$~21 (15 $\mu$m) and 38~$\pm$~10 \% 
(24 $\mu$m for both $AKARI$ and $Spitzer$), respectively.
The line emission in these bands is dominated by \hh\ and ionic lines.
We also estimate the contribution of line emission of the 60 and 90 $\mu$m bands
to be 16~$\pm$~8 and 3~$\pm$~2 \%, respectively,
using $Spitzer$ MIPS SED 52--97 $\mu$m spectra.
The line contributions in these bands are mostly due to [\ion{O}{1}] at 63 $\mu$m.
For the 140 and 160 $\mu$m bands, we expect the line contributions to be less than 5~\%
by scaling the flux of the [\ion{O}{1}] 63 $\mu$m line that we obtain above to
those of [\ion{O}{1}] 145 $\mu$m and [\ion{C}{2}] 153 $\mu$m lines in a shocked region \citep{hol89, all08}.
Figure~\ref{fig_sed} shows the flux distribution of the western shell
excluding the line contributions.

The IR synchrotron flux expected from an extrapolation of the radio fluxes 
applying synchrotron power-law index of $-$0.54 \citep{sha70}
is less than a few percents of the measured IR values,
supporting that the observed continuum is dominated by dust emission.
In order to calculate temperatures of the dusts responsible for the mid- and far-IR emission of Kes~17,
we perform the spectral energy distribution (SED) fits of the flux distribution (Figure~\ref{fig_sed})
with two modified blackbody components.
For the emissivity of the dusts in the SED fits,
we adopt a mixture of carbonaceous and silicate interstellar grains \citep{dra03a}.
As a result,
we obtain acceptable fit (reduced chi-square $<$2) with 
dust temperature components of 79~$\pm$~6~K and 27~$\pm$~3~K (Figure~\ref{fig_sed}).
The mass of the hot (79~K) dust is $(6.2~\pm~4.6)~\times~10^{-4}$~$d_8^2$~$M_\odot$ 
and that of the cold (27~K) dust is 6.7~$\pm$~4.0~$d_8^2$~$M_\odot$.
The luminosities of the dust emission are  (9.5~$\pm$~6.1)~$\times$~10$^2$~$d_8^2$ $L_\odot$ and 
(8.1~$\pm$~5.0)~$\times$~10$^3$~$d_8^2$ $L_\odot$ for the hot and cold components, respectively.
It is worthwhile to note that the hot dust temperature could have been overestimated
if there is any contribution from stochastically heated small grains 
to the 15 and 24 $\mu$m continuum fluxes \citep{dwe86}.

\section{Near-Infrared Spectroscopy and Bright H$_2$ Lines}

\begin{figure}
\plotone{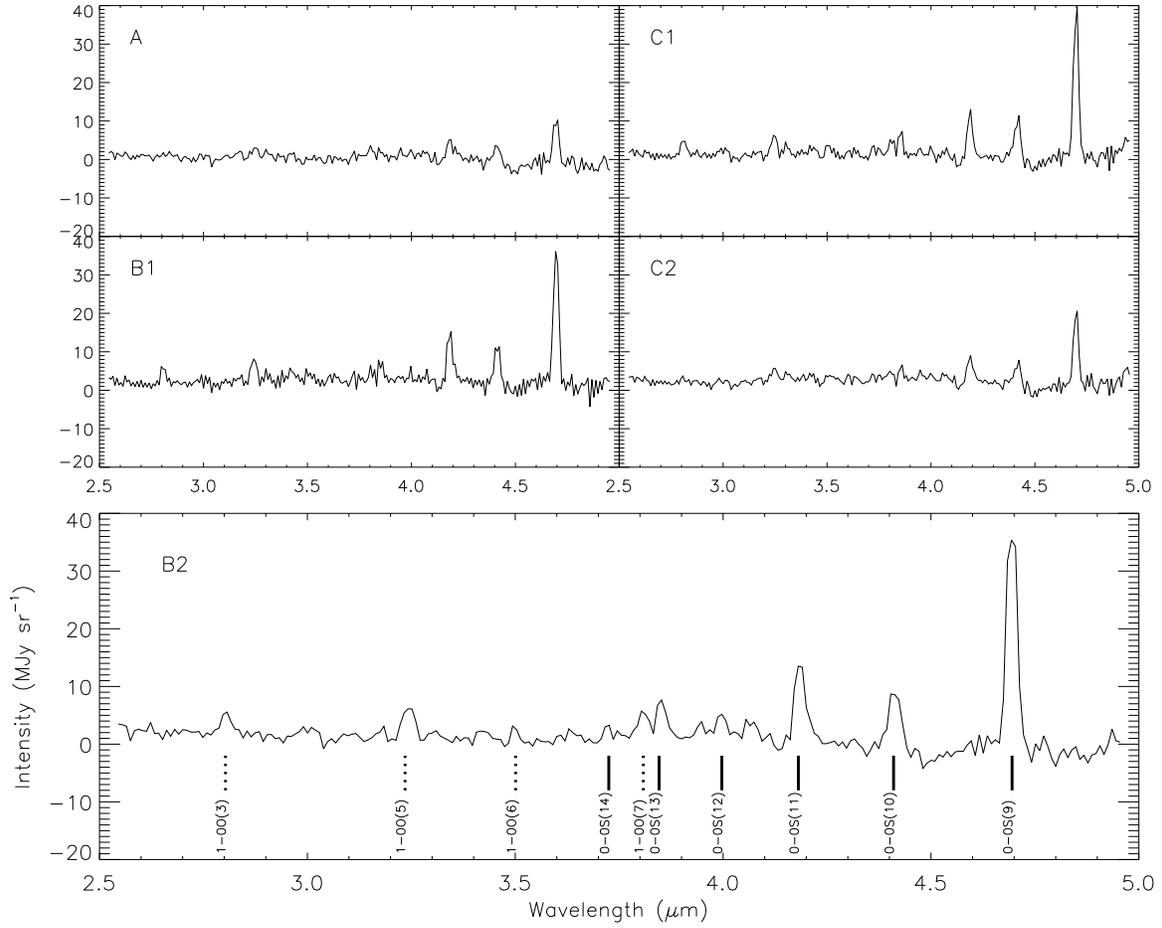}
\caption{The $AKARI$ near-IR spectra of Kes~17 from the five position (A, B1, B2, C1, C2) 
in the western shell. 
Their slit positions and coordinates are listed in 
Figure~\ref{fig_ir} and Table~\ref{tab_slit}, respectively.
We mark the identified pure rotational (solid) and
ro-vibrational (dotted) \hh\ lines in the bottom panel of the B2 position.
}
\label{fig_spec}
\end{figure}

Figure~\ref{fig_spec} shows near-IR (2.5--5.0 $\mu$m) $AKARI$ spectra 
of five points (A, B1, B2, C1, C2 in Figure~1) in the western shell of Kes~17
obtained after subtraction of the background spectrum.
(See Table~4 for their coordinates.)
We detect several \hh\ lines in the spectra including
the pure rotational transitions of 0-0 S(9--14)
and ro-vibrational transitions of 1-0 O(3, 5, 6).
Table~\ref{tab_line} lists the observed intensities of the line emission.

\begin{figure}
\plotone{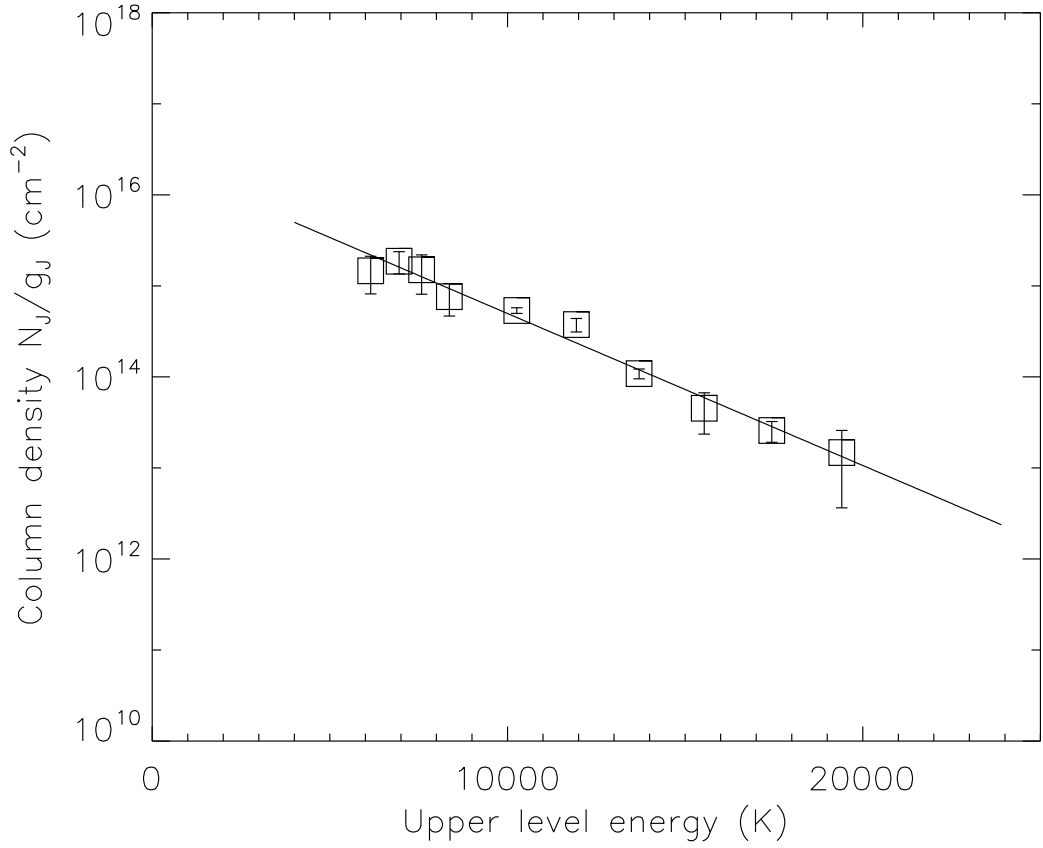}
\caption{The $AKARI$ \hh\ line population diagram of the B2 position.
The solid line shows the results of the linear fit of the observed fluxes
with the excitation temperature of 2,200~K.
}
\label{fig_lpop}
\end{figure}

Figure~\ref{fig_lpop} presents the excitation diagram of the pure rotational and ro-vibrational
\hh\ lines of B2 between the upper energy level of 6,000~K and 20,000~K.
For this we correct the extinction effect using the H column density of 3.6$\times$~10$^{22}$~cm$^{-2}$ 
estimated in previous X-ray observations of Kes~17 and the Galactic extinction curve 
\citep{hew09, dra03a}.
The linear correlation between the upper energy level and the upper column density 
represented by a straight line in Figure~\ref{fig_lpop} indicates the level population 
of the local thermodynamic equilibrium at the excitation temperature of 2200~$\pm$~400~K.
The spectra of other positions, which have smaller number of detected lines, show similar 
excitation temperatures to that of B2.

\section{Discussions}

\subsection{Origin of Bright Infrared Emission of Kes~17}

The IR emission of Kes~17 shows the existence of both the western and southern shell structures.
The former is visible in all the IR bands, while the latter is visible only in the mid-IR bands of 
15 and 24 $\mu$m.
The near-IR emission of the western shell is dominated by \hh\ line emission (Figure~\ref{fig_spec}),
which suggests that Kes~17 is interacting with molecular gas in this region,
consistent with the results of previous observations \citep[e.g.,][]{hew09}.
The distribution of the mid-IR emission overlaps that of the radio continuum emission 
in the western and southern shells (Figure~\ref{fig_mid}),
supporting the interpretation that it represents swept-up dust at the boundary of the SNR.
On the other hand, the far-IR emission is similar to the near-IR emission 
with only the western shell being readily identifiable.
The coexistence of the far-IR emission from the cold dust and
the near-IR line emission form shocked molecular gas
in the western shell suggests that the bright far-IR emission of Kes~17 in this region
is also related to the interaction of the source with a nearby molecular cloud.

The far-IR luminosity of Kes~17 ($\sim$~8100~$d_8^2$~$L_\odot$; \S~3)
makes it one of the most luminous SNRs in the far-IR wavebands 
and the second (after Cassiopeia A) Galactic SNR detected up to 160 $\mu$m 
in the IR wavebands \citep{sib10}.
For comparison, only eight SNRs were detected by IRAS at 100 $\mu$m
with far-IR luminosity larger than Kes~17 \citep{sak92},
five of them interacting with molecular clouds \citep[e.g.,][]{koo97, rea99, keo07, hew09}.
Our observations also strongly suggest that the far-IR emission of Kes~17
is originated from the dust in a molecular cloud interacting with the SNR.
However, molecular shocks producing strong \hh\ lines in a dense molecular cloud
cannot directly generate bright far-IR dust continuum emission. 
This is because, in the non-dissociative molecular shock with shock speeds of $<$ 50 km s$^{-1}$, 
the gas temperature does not rise above a few thousand degrees \citep{dra81}.
This is inadequate to heat the dust grains sufficiently high enough to emit 
bright far-IR continuum emission.
Instead, if a molecular cloud is clumpy, composed of clumps and inter-clump medium, 
then the far-IR emission may be produced by faster shocks propagating 
in the inter-clump medium of lower densities.
Even though the density of inter-clump medium is lower than that of the H$_2$ emitting clumps,
it can still be high enough to contain large amount of dusts 
and reduce shock speeds significantly.
In the following we consider these possibilities.

When SNR shocks propagate into a clumpy molecular cloud, 
hot gas in swept-up inter-clump medium behind shock front collisionally heats
dusts to produce the IR emission.
The estimated dust mass of the western shell of Kes~17 is $6.7~\pm~4.0~d_8^2$~M$_\sun$.
If the cloud initially occupies the entire western region
with normal dust-to-gas ratio of 1~\%,
the pre-shock inter-clump density is $40 \pm 20~d_8^{-1}$ cm$^{-3}$.
Assuming that Kes~17 is in a Sedov phase with radius of $\sim$~7~$d_8$~pc,
the shock velocity in the inter-clump medium is
$v_{ic} \simeq 200~E_{51}^{0.5}~(n_{ic}/40~{\rm cm^{-3}})^{-0.5}~d_8^{-1.5}~\rm km~s^{-1}$,
where $E_{51}$ is the SN explosion energy in the unit of 10$^{51}$ erg
and $n_{ic}$ is the density of the inter-clump medium.
In this swept-up inter-clump gas, the IR surface brightness of the dust 
by strong shocks is \citep{dra81} 
$$
I_{\nu}~\simeq~90~\left(\frac{n_{ic}}{\rm 40~cm^{-3}} \right) 
           \left(\frac{v_{ic}}{\rm 200~km~s^{-1}} \right)^3 
           \left(\frac{\lambda}{\rm 90~\mu m} \right) 
           ~~~ {\rm MJy~sr^{-1}}
$$
around 90 $\mu$m,
which, for the obtained inter-clump density and velocity of $\sim$ 40 cm$^{-3}$ 
and $\sim$ 200 km s$^{-1}$, 
gives comparable surface brightness to the observed value.
Note that the calculation by \citet{dra81} 
expects the peak of dust emission power
($\lambda f_\lambda$) at the mid-IR wavelength ($\sim 30~\mu$m),
while our observations indicate a somewhat flat $\lambda f_\lambda$ peak around
the 65--90~$\mu$m bands. 
This is suggestive that there may be a significantly increased amount of large dusts
contributing effectively in the far-IR regime if the shocked inter-clump medium
is responsible for the observed emission of Kes~17.

Alternatively, the bright far-IR emission of Kes~17 may be the 
emission of dusts injected from an evaporating cloud \citep{dwe81}. 
If so, the expected IR luminosity of an evaporating cloud in 
the hot gas of an SNR is \citep{dwe81}
$$
L_{IR}~\simeq~200~n_{h}^2~
         \left( \frac{R_{c}}{\rm 1~pc} \right)^3  
         \left( \frac{T_h}{\rm 10^{7}~K} \right)^{1.5}  ~~~ L_{\odot}
$$
where $R_{c}$ is the radius of the cloud and $n_{h}$ and $T_h$ are the
density and temperature of the hot gas, respectively.
The X-ray emitting gas was recently detected inside the radio shell of Kes~17 
\citep{com10}, suggesting that the evaporation should occur at the
surface of an IR emitting cloud where it is in contact with the hot interior gas.
The temperature and density of the hot X-ray gas
are $\sim 10^7$~K and $\sim 1$~cm$^{-3}$, respectively \citep{com10}.
We expect the IR luminosity of an evaporating cloud to be $\sim$ 2,000~$d_8^3$ $L_\odot$
using the cloud radius of 3~$d_8$~pc ($\sim1.3'$) at 90 $\mu$m (Figure~\ref{fig_ir}).
This is somewhat smaller than the observed far-IR luminosity ($\sim$ 8000 $L_\odot$, \S~3) of Kes~17, 
indicating that the dust emission from an evaporating cloud 
may contribute only a small amount of the far-IR emission of Kes~17,
although we cannot rule out the possibility completely
given the various uncertainties involved in the measurements.

Another explanation of the bright far-IR emission of Kes~17 may be that
the dusts in a pre-shock molecular cloud heated by strong radiation
of the SNR shocks in a radiative phase emit the observed emission.
The shocks propagating into a molecular cloud at the western shell of Kes~17
can easily become radiative for the pre-shock density of $\simlt$ 100~cm$^{-3}$
if gas cooling is efficient.
Taking a power-law temperature dependence of the cooling efficiencies
\citep[$\propto T^{-0.5}$;][]{kah76},
\citet{cox99} calculated a radiative shell forming at the radius 
$
R_{sh} \simeq 7~E_{51}^{1/8}~t_{10}^{3/4}~ \rm pc 
$
with a shell velocity 
$
v_{sh} \simeq 200~E_{51}^{1/8}~t_{10}^{1/4}~ {\rm km~s}^{-1}
$
for a pre-shock density 
$
n_{o} \simeq 20~E_{51}^{3/8}~t_{10}^{-7/4}~ \rm cm^{-3}
$
and a radiative shell forming time $t_{10}$ = 10,000 yrs.
This indicates that Kes~17 likely has a radiative shell, 
if the pre-shock density is $\simgt$ 20 cm$^{-3}$.
Using the shell forming values and a cloud radius of $R_{c}$ = 3~$d_8$~pc,
the luminosity of a radiative shock at the western shell of Kes~17 is expected to be \citep{hol89}
$$
L \simeq 2.6 \times 10^4 
         \left( \frac{R_{c}}{\rm 3~pc} \right)^2  
         \left( \frac{n_{o}}{\rm 20~cm^{-3}} \right)  
         \left( \frac{v_{s}}{\rm 200~km~s^{-1}} \right)^3 ~~~ L_\odot 
$$
which appears to be adequate to explain the observed luminosity of the western shell (\S~3).
In this case, we expect the peak of the dust emission power
to be at the far-IR ($\ge$ 60 $\mu$m) region
for the dusts of mixture of carbonaceous and silicate grains \citep{dra11},
consistent with the results of this study.
This can also explain the absence of the X-ray emission in the western shell of Kes~17
as the lack of X-ray emitting high temperature gas in the radiative phase \citep{com10}.

\subsection{Thermal Admixture Model of Molecular Shocks}

Our near-IR (2.5--5.0 $\mu$m) spectra of the western shell of Kes~17
indicate the excitation temperature of 2200~$\pm$~400~K for \hh\ gas
in this region (Figure~\ref{fig_spec}).
This is different from the previous results based on the $Spitzer$ observations
where a mixture of two components of \hh\ gas of $\sim$ 300~K and $\sim$ 1200~K
was proposed to explain the observed intensity distribution of \hh\ lines of the western shell
in the 5--28 $\mu$m range \citep{hew09}. 
The discrepancy in the excitation temperature suggests that the 
\hh\ gas in the western shell is in a thermal admixture
as found in other \hh\ gas interacting with SNRs \citep{neu08, shi09, shi10}.

\begin{figure}
\plotone{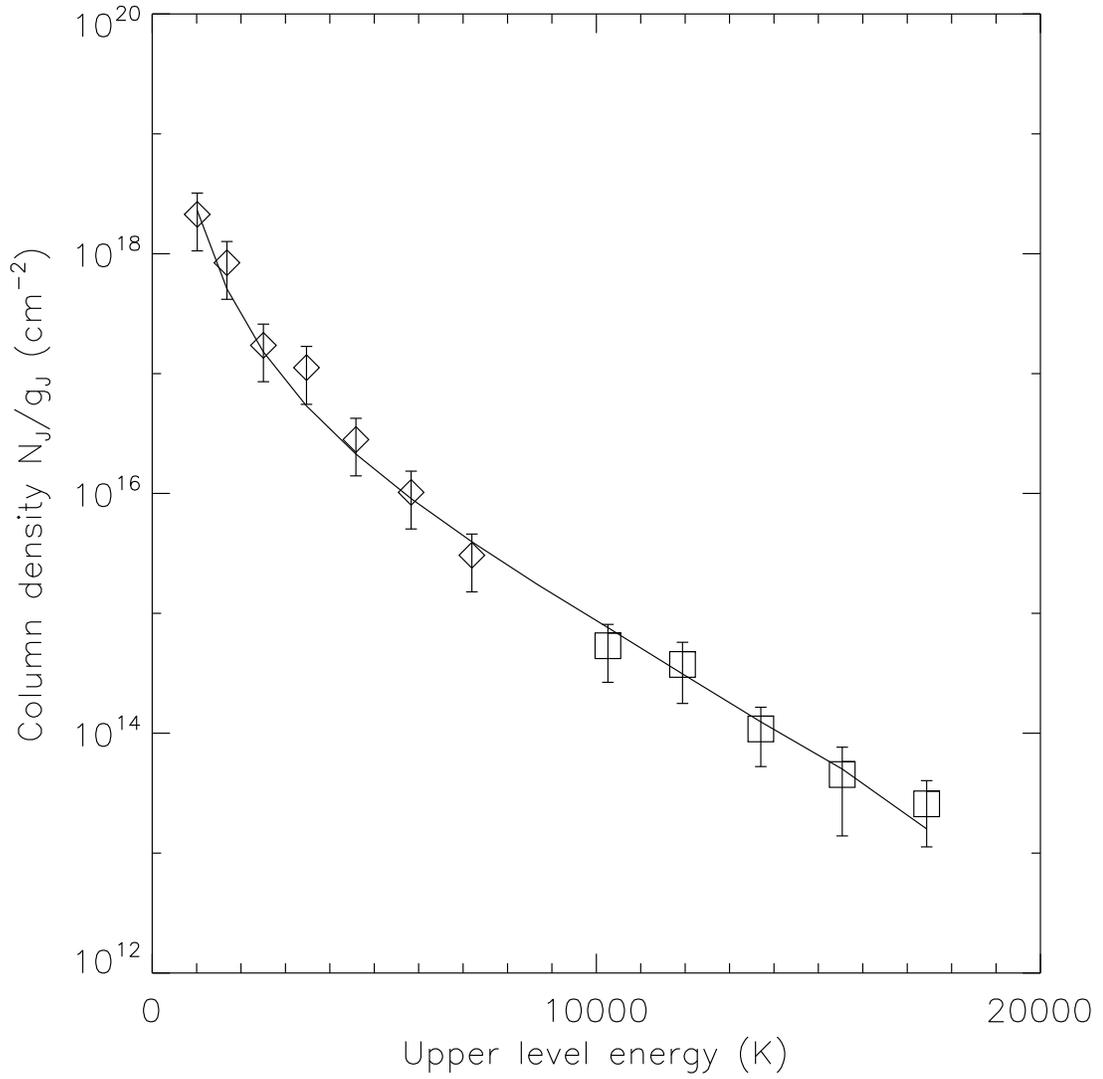}
\caption{Combined $AKARI$ (square) and $Spitzer$ (diamond)
\hh\ line population diagram of the western shell.
The solid curve represents the prediction of the best-fit 
parameters of the thermal admixture model.
}
\label{fig_lcal}
\end{figure}

Figure~\ref{fig_lcal} shows the excitation diagram of \hh\ lines
detected in both the $Spitzer$ and $AKARI$ observations,
covering the upper energy level of 1000--17,000~K.
(Note that while the $Spitzer$ results represent the levels of $\simlt$ 7,000~K, 
the $AKARI$ results do the levels of higher temperatures.)
We fit the observed excitation diagram with a thermal admixture model
where the column density of \hh\ gas is related to a power of its temperature of 100--4,000~K:
$dN \sim T^{-b} dT$ \citep{shi09, shi10}.
The ortho-to-para ratio is fixed to be 3 in the fit.
The best fit gives 3.0~$\pm$~0.1, ($1.5~\pm~0.2)~\times~10^{21}$~cm$^{-2}$, 
and ($3.2~\pm~0.7)~\times~10^5$~cm$^{-3}$ 
for the power index, column density, and number density, respectively.
The solid curve in Figure~\ref{fig_lcal} represents the best-fit model
to the observed data, where we can identify that the model prediction
matches the observed values well.
This column density is twice that of the two temperature model of
the $Spitzer$ observations, while the number density is between 
those obtained for the warm and hot components of the $Spitzer$ results \citep{hew09}.
The obtained power index of 3.0~$\pm$~0.1 is comparable to what obtained
from other sources \citep{shi09, neu09}, and is not far from 3.8 obtained 
for bow shocks \citep{neu08}.
If Kes~17 is indeed interacting with clumpy molecular gas (see \S~5.1),
it can be easy to develop a group of shocks around dense clumps with 
various shock velocities,
which in turn naturally leads to an admixture of \hh\ gas of different temperatures.

\section{Conclusion}

We show in this paper that Kes~17 is one of the most luminous SNRs in the IR wavebands,
reaching the far-IR luminosity of $\sim$ 8100 $L_\odot$.
Its IR emission is concentrated on the western and southern shell structures.
The western shell is bright in the mid- and far-IR continuum and the near-IR \hh\ line emission,
whereas the southern shell is visible only in the mid-IR continuum emission.
The far-IR continuum and near-IR \hh\ line emission of the western shell is related
to its interaction with nearby molecular gas.
It is apparent that there exists dense shocked molecular gas, excited to $\sim$ 2,000~K, 
in this region that produces the observed \hh\ line emission.
If the molecular gas is clumpy, the observed far-IR emission can be produced
by the dust associated with inter-clump molecular gas of lower densities.
The far-IR emission may originate from
the dust emission of the swept-up inter-clump molecular gas,
the dust emission from an evaporating cloud,
or the dust emission exposed to strong radiative shocks.
The mid-IR emission of Kes~17 is bright in both the western and
southern shells and overlaps radio continuum emission
which forms a partly-broken circular shell structure.
This suggests that the mid-IR and radio continuum emission delineates
the boundary of Kes~17 made by the swept-up material.

The SED distribution of the IR emission of Kes~17 indicates the existence of
two dust components: hot (79~$\pm$~6~K) component of $(6.2~\pm~4.6)~\times~10^{-4}$~$d_8^2$~$M_\odot$
and cold (27~$\pm$~3~K) component of 6.7~$\pm$~4.0~$d_8^2$~$M_\odot$.
The former is responsible for most of the mid-IR continuum emission,
while the latter is for the far-IR emission.
The observed flux distribution of the \hh\ lines, on the other hand,
can be explained with a thermal admixture of the shocked molecular gas of
different temperatures in the 100--4,000~K range.
This thermal admixture of the shocked molecular gas may be a natural
consequence of the interactions between Kes~17 and clumpy molecular gas
in which various shock velocities can easily produce thermal admixture 
of the molecular gas.
Overall, it appears that combined far-IR continuum and 
near-IR spectroscopic observations are very useful to studying
SNRs interacting with molecular gas, as was the case for Kes~17 in this paper,
providing information 
for both the gas and dust components associated with the shock interactions.

This work is based on observations with $AKARI$, 
a JAXA project with the participation of ESA.
This work is based in part on observations 
made with the $Spitzer$ Space Telescope, 
which is operated by the Jet Propulsion Laboratory, 
California Institute of Technology under a contract with NASA.
The Australia Telescope Compact Array 
is part of the Australia Telescope 
which is funded by the Commonwealth of 
Australia for operation as a National Facility managed by CSIRO.
We thank all the members of the $AKARI$ project.
We also thank an anonymous referee for constructive comments.
H.-G.L. was supported by the Early Research Award Program (ERA07-03-270)
to D.-S.M. from Ministry of Research and Innovation of the Ontario Provincial Government.
D.-S.M. acknowledges the support by NSERC through Discovery program 327277.
B.-C.K. was supported by the National Research Foundation of Korea (NRF)
Grant (NRF-2010-616-C00020).

{\it Facility:} \facility{$AKARI$, $Spitzer$, $ATCA$}

{}

\clearpage

\clearpage

\clearpage

\clearpage

\clearpage

\clearpage
\clearpage

\clearpage
\begin{deluxetable}{l ccc}
\tabletypesize{\scriptsize}
\tablewidth{0pt}
\tablecolumns{4}
\tablecaption{Summary of $AKARI$ observations
\label{tab_obssum}}
\tablehead {
\colhead{Date} &\colhead{ID\tablenotemark{a}} &\colhead{Mode} &\colhead{Band}
}
\startdata
2007. 2. 5.       & 1400755     & IRC imaging     & L15, L24 \\
2007. 2. 6.       & 1400754     & FIS imaging     & N60, Wide-S, Wide-L, N160 \\
2008. 8. 8.       & 1420785(1,2), 1420783(1,2), 1421783 & IRC spectroscopy & NG \\
2009. 2. 4.       & 1420784     & IRC spectroscopy & NG \\
2009. 2. 6.       & 1421784, 1421785 & IRC spectroscopy & NG \\
\enddata
\tablenotetext{a}{$AKARI$ observational identification number.
}
\end{deluxetable}

\begin{deluxetable}{l ccccc cccc}
\tabletypesize{\tiny}
\tablewidth{0pt}
\tablecolumns{10}
\tablecaption{Characteristics of the imaging bands.\tablenotemark{a}
\label{tab_obsimg}}
\tablehead {
\colhead{Band} 
& \colhead{$Spitzer$} & \colhead{$Spitzer$}& \colhead{$AKARI$} & \colhead{$Spitzer$} & \colhead{$AKARI$}
& \colhead{$AKARI$} & \colhead{$AKARI$} & \colhead{$AKARI$} & \colhead{$AKARI$} 
\\
& \colhead{4.5 $\mu$m} & \colhead{8 $\mu$m}& \colhead{15 $\mu$m} & \colhead{24 $\mu$m} & \colhead{24 $\mu$m}
& \colhead{65 $\mu$m} & \colhead{90 $\mu$m} & \colhead{140 $\mu$m} & \colhead{160 $\mu$m} 
}
\startdata
\colhead{Instrument}
& IRAC~4.5 &IRAC~8 &IRC~L15 &MIPS~24 &IRC~L24
& FIS N60 & FIS Wide-S & FIS Wide-L & FIS N160 \\
Reference wavelength ($\mu$m)
& 4.49 & 7.87 & 15.0 & 23.68 & 24.0
& 65  & 90  & 140 & 160 \\
Effective bandwidth  ($\mu$m)
& 1.01 & 2.93 & 5.98 & 5.3 & 5.34
& 21.7& 37.9& 52.4& 34.1\\
FWHM                 ($''$)
& 1.72 & 1.98 & 5.7 & 6 & 6.8
& 37  & 39  & 58  & 61
\enddata
\tablenotetext{a}{Detailed characteristics of 
$AKARI$ IRC, $AKARI$ FIS $Spizter$ IRAC, and $Spizter$ MIPS 
are described in 
\citet{ona07}, \citet{kaw07}, \citet{faz04}, and \citet{rie04},
respectively.}
\end{deluxetable}

\begin{deluxetable}{l cc}
\tabletypesize{\scriptsize}
\tablewidth{0pt}
\tablecolumns{3}
\tablecaption{Slit positions of $AKARI$ spectroscopic observations.
\label{tab_slit}}
\tablehead {
\colhead{Slit} &\colhead{Position\tablenotemark{a}} &\colhead{Exposure}
}
\startdata
A 
&($\rm 13^h05^m48^s$, $\rm -62^\circ38'39''$)
& 1 $\times$ 308 s
\\
B1 
&($\rm 13^h05^m33^s$, $\rm -62^\circ40'00''$)
& 2 $\times$ 308 s
\\
B2 
&($\rm 13^h05^m33^s$, $\rm -62^\circ40'19''$)
& 1 $\times$ 308 s
\\
C1
&($\rm 13^h05^m32^s$, $\rm -62^\circ42'34''$)
& 1 $\times$ 308 s
\\
C2
&($\rm 13^h05^m33^s$, $\rm -62^\circ42'16''$)
& 1 $\times$ 308 s
\\
Background\tablenotemark{b}
&($\rm 13^h05^m20^s$, $\rm -62^\circ39'11''$)
& 1 $\times$ 308 s
\\
\enddata
\tablenotetext{a}{
The slit position angles were fixed to be 127$^\circ$
by the satellite orbit.
}
\tablenotetext{b}{
One of the accompanying $3''\times~1.0'$ slit data
was used for the background spectrum.
}
\end{deluxetable}

\begin{deluxetable}{lccc}
\tabletypesize{\scriptsize}
\tablewidth{0pt}
\tablecolumns{4}
\tablecaption{Flux of the western and southern shells.
\label{tab_flux}}
\tablehead {
\colhead{Band}     &\multicolumn{2}{c}{Western shell}              &\colhead{Southern shell} 
\\
\colhead{}         &\colhead{Flux\tablenotemark{a}} &\colhead{Line contribution\tablenotemark{b}}
&\colhead{Flux\tablenotemark{a}}          
}
\startdata
$AKARI$ 15 $\mu$m          & 2.6 $\pm$ 0.5 Jy &  75  $\pm$ 21  \%               &  1.2 $\pm$ 0.2 Jy \\
$Spitzer$ 24 $\mu$m        & 2.2 $\pm$ 0.4 Jy &  38  $\pm$ 10  \%               &  1.5 $\pm$ 0.2 Jy \\
$AKARI$ 24 $\mu$m          & 3.1 $\pm$ 0.7 Jy &  39  $\pm$ 11  \%               &  1.0 $\pm$ 0.3 Jy \\
$AKARI$ 65 $\mu$m          & 75  $\pm$  15 Jy &  16  $\pm$ 8  \%                &  $<$4  Jy \\
$AKARI$ 90 $\mu$m          & 90  $\pm$  27 Jy &  3   $\pm$ 2  \%                &  $<$8  Jy \\
$AKARI$ 140 $\mu$m         & 104 $\pm$  44 Jy &  $<$5 \%                               &   $<$53 Jy \\
$AKARI$ 160 $\mu$m         & 75  $\pm$  31 Jy &  $<$5 \%                               &   $<$34 Jy \\
\enddata
\tablenotetext{a}{
Measured total band flux including continuum and line emission.
}
\tablenotetext{b}{
We estimate the line contribution to the observed band flux 
using the $Spitzer$ IRS and MIPS SED spectra
}
\end{deluxetable}

\begin{deluxetable}{lc ccccc}
\tabletypesize{\scriptsize}
\tablewidth{0pt}
\tablecolumns{7}
\tablecaption{Surface brightness of the observed \hh\ lines.
\label{tab_line}}
\tablehead {
\colhead{Line} &\colhead{Wavelength} 
                       &\colhead{A} &\colhead{B1} &\colhead{B2} &\colhead{C1} &\colhead{C2}
\\
\colhead{}     &\colhead{($\mu$m)} 
                       & \multicolumn{5}{c}{(10$^{-5}$ erg cm$^{-2}$ s$^{-1}$ sr$^{-1}$)}
}
\startdata
H$_2$ 1-0 O(3)  & 2.80 &            & 11.6 (4.8)  & 9.7 (4.3)   & 12.4 (4.7)  & 2.8 (2.0)   \\
H$_2$ 1-0 O(5)  & 3.24 & 4.3 (3.1)  & 10.6 (6.6)  & 16.4 (4.5)  & 10.3 (3.1)  & 5.7 (2.9)   \\
H$_2$ 1-0 O(6)  & 3.50 &            &             & 4.2 (2.0)   & 7.8 (4.0)   &             \\
H$_2$ 0-0 S(14) & 3.72 &            &             & 2.5 (1.9)   & 2.9 (1.7)   &             \\
H$_2$ 1-0 O(7)  & 3.81 &            & 2.6 (1.4)   & 5.8 (2.2)   & 6.1 (3.3)   & 2.2 (2.3)   \\
H$_2$ 0-0 S(13) & 3.85 &            & 6.9 (2.3)   & 8.4 (2.2)   & 7.7 (2.4)   & 3.5 (1.4)   \\
H$_2$ 0-0 S(12) & 4.00 &            &             & 3.6 (1.7)   & 1.7 (1.4)   &             \\
H$_2$ 0-0 S(11) & 4.19 & 7.1 (2.5)  & 16.0 (2.7)  & 18.6 (2.3)  & 17.1 (2.1)  & 10.7 (2.2)  \\
H$_2$ 0-0 S(10) & 4.41 & 8.5 (2.6)  & 14.0 (2.4)  & 14.5 (2.5)  & 15.3 (2.5)  & 7.8 (2.1)   \\
H$_2$ 0-0 S(9)  & 4.69 & 13.1 (3.7) & 35.1 (1.9)  & 39.5 (2.9)  & 37.5 (1.9)  & 18.5 (1.7)  \\
\enddata
\tablecomments{
The numbers in parentheses are measured errors
}
\end{deluxetable}

\end{document}